\documentclass[12pt,a4paper]{article}

\usepackage{amssymb,amsfonts}
\usepackage[centertags]{amsmath}
\usepackage{graphicx,psfrag}

\newcommand{\be}{\begin{equation}}
\newcommand{\ee}{\end{equation}}
\newcommand{\bea}{\begin{eqnarray}}
\newcommand{\eea}{\end{eqnarray}}
\newcommand{\bdm}{\begin{displaymath}}
\newcommand{\edm}{\end{displaymath}}
\newcommand{\lb}{\label}

\newcommand{\I}{\mbox{i}}
\newcommand{\D}{\mbox{d}}
\newcommand{\E}{\mbox{e}}

\newcommand{\bvec}[1]{\mathbf{#1}}
\newcommand{\cH}{\mathcal H}
\newcommand{\abs}[1]{\left\lvert#1\right\rvert}

\begin{document}
\begin{titlepage}
\begin{center}
{\large\bf  POINTER STATES FOR PRIMORDIAL FLUCTUATIONS\\
                      IN INFLATIONARY COSMOLOGY}
\vskip 1cm
{\bf Claus Kiefer and Ingo Lohmar}
\vskip 0.4cm
 Institut f\"ur Theoretische Physik, Universit\"at zu K\"oln,\\
 Z\"ulpicher Str.~77, 50937 K\"oln, Germany.
\vskip 0.7cm
{\bf David Polarski}
\vskip 0.4cm
 Laboratoire de Physique Th\'eorique et Astroparticules, UMR 5207 CNRS,\\
  Universit\'e de Montpellier II, 34095 Montpellier, France.
\vskip 0.7cm
{\bf Alexei A. Starobinsky}
\vskip 0.4cm
 Landau Institute for Theoretical Physics,
 Kosygina St. 2, Moscow 119334, Russia.

\end{center}
\date{\today}
\vskip 2cm
\begin{center}
{\bf Abstract}
\end{center}

\small
\begin{quote}
Primordial fluctuations in inflationary cosmology acquire classical 
properties through decoherence when their wavelengths become larger 
than the Hubble scale. Although decoherence is effective, it is not 
complete, so a significant part of primordial correlations remains up 
to the present moment. We address the issue of the pointer states 
which provide a classical basis for the fluctuations with respect to 
the influence by an environment (other fields). Applying methods from 
the quantum theory of open systems (the Lindblad equation), we show 
that this basis is given by narrow Gaussians that approximate 
eigenstates of field amplitudes. We calculate both the von Neumann and 
linear entropy of the fluctuations. Their ratio to the maximal entropy 
per field mode defines a degree of partial decoherence in the entropy 
sense. We also determine the time of partial decoherence making the 
Wigner function positive everywhere which, for super-Hubble modes 
during inflation, is virtually independent of coupling to the 
environment and is only slightly larger than the Hubble time. On the 
other hand, assuming a representative environment (a photon bath), the
decoherence time  
for sub-Hubble modes is finite only if some real dissipation exists.
\end{quote}
\normalsize
\end{titlepage}


\section{Introduction}

According to the inflationary scenario, all structure in the
Universe originates from quantum vacuum fluctuations during a de
Sitter (inflationary) stage in the very early Universe. These are
inhomogeneous fluctuations (perturbations) of both the space-time
metric and a scalar inflaton field. While the tensor part of
the metric fluctuations generated during inflation produces the
primordial gravitational wave background \cite{St79}, its scalar
part together with perturbations of the inflaton leads to the
origin of primordial density fluctuations producing present
gravitationally bound objects and the large-scale structure of the
Universe \cite{MC81}. All these fluctuations can be formally
represented by a scalar field with a time-dependent `mass', the
time dependence coming from the coupling to the expanding universe
described by a scale factor $a(t)$ (we assume here a spatially
flat Friedmann-Robertson-Walker (FRW) cosmological model).

Usually one assumes these fluctuations to be in their ground state
(the adiabatic vacuum state) at the onset of inflation. This
choice follows from the hypothesis of the maximal possible
symmetry of the Universe in some period in the past during
inflation. Also, it can be shown that this initial condition for
fluctuations is an attractor for a wide open set of other initial
conditions with a non-zero measure. The smallness of the
fluctuations means that different modes (distinguished by a wave
vector ${\mathbf k}$ with the wave number $k=|{\mathbf k}|$) decouple
and can be treated separately. Modes relevant for structure
formation cross the Hubble radius $H^{-1}$ twice. Here
$H\equiv\dot{a}/a$ is the Hubble parameter (with the dot denoting
$d/dt$). During inflation,
their wavelength $\lambda=2\pi a/k$ becomes bigger than $H^{-1}$
(`the first Hubble radius crossing'). The quantum modes then acquire
classical properties in the following sense. First, even without
considering any interaction with other degrees of freedom,
expectation values of any physical quantities constructed from the
quantum fluctuations become practically indistinguishable from
mean values of corresponding classical quantities as functions of
classical stochastic fluctuations. This is achieved due to huge
squeezing (in the sense of quantum optics) of those modes that
gives a possibility to neglect non-commuting parts of all mode
quantum operators. Since no consideration of environmental degrees
of freedom and their interaction with modes involved are needed
for this indistinguishability, this quantum-to-classical
transition was called `decoherence without decoherence' \cite{PS1}. 
This result can be generalized to all higher non-linear orders 
of metric perturbations if only the growing (quasi-isotropic) mode of 
perturbations is kept \cite{LS06}. In this way, it is possible to 
explain the classical stochastic behaviour of observed cosmological 
density fluctuations including their power spectrum and statistics. 
However, more subtle questions like the correct calculation of the 
{\em entropy} of primordial fluctuations cannot be solved in this 
approximation.

Second, this classical behaviour is preserved, and even
re-enforced, in the presence of an `environment'. Since such an
environment is always present in the form of other (`irrelevant')
fields and fluctuations (or non-linear couplings of the same
field), its influence {\em has} to be taken into account. In
laboratory, it is usually the environment that leads to the
emergence of classical behaviour, a process which is called
decoherence \cite{deco}. In particular, strongly squeezed quantum
states are known to be especially sensitive to decoherence. We
have demonstrated some time ago that the classical basis
distinguished by the environment (the `pointer basis') is -- in
the limit of large squeezing -- approximately given by the basis
of {\em field amplitudes} \cite{KPS1} (see also \cite{KP} for a
detailed conceptual discussion). The correct choice of the pointer
basis is of particular relevance for the entropy of primordial
fluctuations. We have calculated the entropy for each mode and
found that it can assume at most half of its maximal value before
the second Hubble radius crossing \cite{KPS00}. The maximal
entropy would be achieved if the pointer basis was the
particle-number basis \cite{prokopec}.

The wavelength of the modes becomes again smaller than $H^{-1}$
(`the second Hubble radius crossing') during the radiation and
matter dominated phases. The scalar modes have left their imprint
on the anisotropy spectrum of the cosmic background radiation,
where they can be observed in the form of acoustic peaks (see e.g.
\cite{Archeops,WMAP} for recent observations). Oscillations of the
same origin (the `Sakharov oscillations', often called baryon acoustic
oscillations) have been recently
discovered in the three dimensional power spectrum of galaxy 
inhomogeneities \cite{E05}. These peaks would be absent if the 
entropy of each mode assumed its maximal value \cite{KPS00}. This 
demonstrates that questions of the quantum-to-classical transition for
primordial perturbations are not only of academic nature, but have
observational relevance. Primordial gravitational waves (tensor
modes) generated during inflation also produce primordial peaks in
the CMB temperature anisotropy and polarization multipoles (see
\cite{JPPS00} for discussion of perspectives of their
observation).

In spite of these investigations, the discussion about the precise
mechanism of this quantum-to-classical transition and the amount
of entropy of primordial fluctuations gained in its course goes
on, cf. the recent papers \cite{CP1,LLN,Hogan,CP2,Martineau,BHH}.
It was argued, for example, that the pointer basis is not given by
the field-amplitude basis, but by the basis of two-mode coherent
states \cite{CP1}, by the squeezed-state basis \cite{Martineau},
or by the second-order adiabatic basis \cite{amm05}.
It was further argued in \cite{CP1} that the associated entropy of
fluctuations (being half the maximal entropy) provides a lower
(instead of upper) bound on the entropy. Also, there is a
disagreement between \cite{Martineau} and \cite{BHH} regarding
the question whether 
decoherence may occur already during inflation (after the first
Hubble radius crossing, of course) or only after its end, during
reheating. We find it therefore appropriate to address this issue
again by reviewing and extending our line of arguments.

Our article is organized as follows. 
In Section~2 we first state the problem of the quantum-to-classical
transition and briefly review the central general concepts. We then
apply these concepts to the primordial fluctuations.  
We give exact and
approximate expressions for the von Neumann entropy and also
calculate the linear entropy. We then discuss why the pointer states
for the fluctuations is given by approximate field-amplitude
eigenstates (narrow Gaussians)
and not by another basis such as the coherent-state basis.
In Section~3 we apply methods from open-system
quantum theory (the Lindblad equation) to calculate in a general
setting decoherence
times for modes outside and inside the Hubble radius.
We conclude from this again that the pointer states are
narrow Gaussians in the field amplitude. The entropy per mode is
thus bounded from above by half the maximal entropy in the
long-wave regime before the second Hubble radius crossing. 


\section{Quantum-to-classical transition and entropy}

\subsection{Quantum description for the primordial fluctuations} 

The primordial fluctuations which lead to the observed structure in
the Universe can originate from the quantum fluctuations of the metric
and a scalar field in an early phase where the Universe is expanding
in a (quasi-) exponential way (`inflation'). These fluctuations are
assumed to be small so that their mutual interaction can be neglected
in the description of the main scenario. 
This linear treatment of the fluctuations is certainly justified
by the smallness of the observed CMB anisotropy
and of the ratio of the modulus of the Newtonian potential to the square of
the light velocity for all observed gravitationally bound structures in the
Universe (apart from the very close vicinity of black holes).
It is thus appropriate to   
 represent the modes in Fourier spaces where the various modes (wave
 numbers) decouple form each other. The dynamics of both the tensor
 and the scalar modes can be represented by some scalar field
$\phi_{\mathbf k}$. Because of the mode decoupling
we shall in the following skip the index ${\mathbf
k}$. It turns out to be convenient to work with the rescaled field amplitude
$y\equiv a\phi$. In the case of tensor fluctuations (gravitational
waves), this is already the appropriate variable to deal with; for
scalar perturbations one has to use a gauge-invariant combination
of metric and inflaton perturbations. These details are irrelevant for
the study of the quantum-to-classical transition.

Since we are working in Fourier space, $y$ is complex. 
This reflects the fact that we shall have a two-modes state, the 
two modes being given by ${\mathbf k}$ and $-{\mathbf k}$. 
For the following discussion it is, however, sufficient to assume that
$y$ represents one real mode, but one should keep in mind that
the actual number of degrees of freedom is twice as much.
This also holds for the entropy discussed below; the entropy for 
the two-mode state is twice as much as for the one mode considered here.

The dynamics is then described by the Hamilton operator
\be
\lb{H}
\hat{H}=\frac12\left(p^2+k^2y^2+\frac{2a'}{a}yp\right)\ ,
\ee
where $a'\equiv \D a/\D\eta$, with $\eta$ denoting conformal time,
and $p$ is the momentum canonically conjugate to $y$. (We have skipped
the index $k$.) This Hamiltonian follows after expanding the metric
and the scalar field up to second order in the inhomogeneities, with
the cosmological background given by a flat Friedmann-Lema\^{\i}tre
universe. 

Each mode satisfies the Schr\"odinger equation
\be
{\rm i}\hbar\frac{\partial\psi(y,\eta)}{\partial\eta}=\hat{H}\psi(y,\eta)\ .
\ee
Assuming that the
modes are initially (at the onset of inflation) in their ground state,
their wave function is then at any time of the Gaussian form,
\be
\lb{psi}
\psi(y,\eta)=\left(\frac{2\Omega_{\rm R}(\eta)}{\pi}\right)^{1/4}
\exp\left(-\Omega(\eta)y^2\right)\ , \quad
\Omega\equiv\Omega_{\rm R}+\I\Omega_{\rm I}\ .
\ee
Non-Gaussian initial states can also be studied \cite{LPS};
it was shown there that there is a wide class of initial
non-vacuum states that cannot be distinguished from a vacuum initial
state by just looking at the statistics of observable quantities.

One can prove from the form of the Hamiltonian (\ref{H}) that the
states \eqref{psi} represent squeezed states; the squeezing is
generated by the last term in \eqref{H}. (Actually, we have a
two-modes squeezed state, but for simplicity we present one mode
only, cf. the remark above.)
Introducing the squeezing
parameter $r$ and the squeezing angle $\varphi$, one can write
(setting $\hbar=1$ from now on)
\be \lb{Omega} \Omega_{\rm R}=\frac{k}{\cosh
2r+\cos2\varphi\sinh 2r}\ , \quad \Omega_{\rm I}=-\Omega_{\rm
R}\sin2\varphi\sinh2r \ ,
\ee
with $\Omega_{\rm R}=k$ and $\Omega_{\rm I}=0$ for the initial
vacuum state. Both $r$ and $\varphi$ are functions of $t$ (resp. $\eta$), 
the exact
dependence arising from the expansion law $a(t)$. For pure
exponential inflation described by 
\bdm
a(t)=a_0\exp(H_{\rm I}t)=-\frac{1}{(\eta-2\eta_{\rm e})H_{\rm I}}\ , 
\edm
where $\eta_{\rm e}$ denotes the conformal time at the end of inflation,
simple
analytic results are available corresponding to the adiabatic vacuum
in the de~Sitter space-time \cite{PS1,AFJP}:
\be
\sinh r=\frac{aH_{\rm I}}{2k}\ , \quad\cos2\varphi=\tanh r \ .
\ee
This would lead to $r\to\infty$, $\varphi\to 0$ if inflation lasted
arbitrarily long. Inflation, however, ends, but it still leads to
$r\approx 120$ for the largest cosmological states \cite{GS}, so one
can certainly neglect after some time terms of the order of
$\exp(-r)$ (`decaying mode'). From (\ref{Omega}) one then finds the
asymptotic values \be \Omega_{\rm R}\to k\E^{-2r}\ , \quad
\Omega_{\rm I}\to -k\E^{-r} \ , \ee 
exhibiting $\Omega_{\rm I}\gg \Omega_{\rm R}$ (WKB limit).
This behaviour is assumed to
occur generically for an inflationary model. One recognizes that
the width of the Gaussian (\ref{psi}) becomes very broad in $y$
and highly squeezed in $p$. This broadness reflects the fact that
the kinetic term of the Hamiltonian becomes negligible in this
limit; in the Heisenberg picture this is exhibited by the fact
that the kinetic term of the Hamiltonian, $\hat{H}_{\rm kin}$,
then approximately commutes with the field-amplitude operator,
$\hat{y}$. In the limit of high squeezing, quantum expectation
values become indistinguishable from mean values of classical
stochastic quantities, the difference containing typically terms
of the order $\exp(-r)$ or powers thereof. 
In a quasi de Sitter stage, the annihilation operator evolves according
to
\be
a(\mathbf k) \to \frac{aH_{\rm I}}{2k}\left( a(\mathbf k)_0 -
              a^{\dagger}(-\mathbf k)_0 \right)
\ee
for $k\ll aH$.
So far, the primordial fluctuations behave fully quantum. They are
described by states which are very broad wave packets. Such broad
packets would exhibit in a laboratory situation very non-classical
behaviour, as could be checked in interference experiments. The
standard scenario of structure formation starts, however, from
classical stochastic fluctuations. A major issue is thus to describe
the
quantum-to-classical transition for the primordial modes. We shall now
first briefly review the general concepts and subsequently shall apply them
to understand the emergence of classical behaviour for these modes.

\subsection{ General concepts of the quantum-to-classical transition}

The superposition principle is at the heart of quantum theory. As a
consequence, the set of classical states is of measure zero, since one
can always superpose different classical states (e.g. different narrow
wave packets) to get `weird' non-classical states. 

Such states may occur in many ordinary measurement situations. 
If we assume that a
measured system is initially in the state $|n\rangle$ and the 
measurement device in
some initial state $|\Phi_0\rangle$,
the evolution according to the Schr\"odinger equation
in the simplest case reads
\be |n\rangle|\Phi_0\rangle \stackrel{t}{\longrightarrow}
     \exp\left(-{\rm i} H_{\rm int}t\right)|n\rangle|\Phi_0\rangle
     =|n\rangle|\Phi_n(t)\rangle\ ,  \label{ideal} \ee
where $H_{\rm int}$ is a special interaction Hamiltonian 
(assumed here to dominate over the free Hamiltonians) which
correlates the system state with the device without changing the former.
The resulting apparatus states $|\Phi_n(t)\rangle$ are often called
`pointer states'. In order to yield distinguishable outcomes, these
pointer states must be approximately orthogonal. 
A process analogous to (\ref{ideal}) can also be
formulated in classical physics. The essential new quantum features
now come into play when one considers a {\em superposition} of different
eigenstates (of the measured `observable') as the initial state. The
linearity of time evolution immediately leads to
\be \left(\sum_n c_n|n\rangle\right)|\Phi_0\rangle
    \stackrel{t}\longrightarrow\sum_n c_n|n\rangle
    |\Phi_n(t)\rangle\ . \label{measurement}
\ee
But this state is a superposition of macroscopic measurement results.
It is by now well established, both theoretically and experimentally,
that the ubiquitous and unavoidable interaction with the environment
has to be taken into account \cite{deco}. 
The measurement device is itself `measured'
(passively recognized) by the environment according to
\be \left( \sum_n c_n|n\rangle|\Phi_n\rangle\right)|E_0 \rangle
  \quad  \stackrel{t}\longrightarrow \quad
   \sum_n c_n|n\rangle |\Phi_n\rangle |E_n\rangle .\label{ideal2}
\ee
This is again a macroscopic superposition, now including the
myriads of degrees of freedom pertaining to the environment
(gas molecules, photons, etc.). However, most of these
environmental degrees of freedom are inaccessible.
Therefore, they have to be integrated out from the full state
(\ref{ideal2}). This leads to the reduced density matrix
for system plus apparatus, which contains all the information
that is available there. It reads
\be
\rho_{\rm SA} \approx \sum_n |c_n|^2 |n\rangle\langle n|
   \otimes |\Phi_n \rangle \langle \Phi_n |
   \qquad\mbox{if}\qquad
   \langle E_n|E_m \rangle \approx \delta_{nm}\ ,
 \label{deco}
\ee
since under realistic conditions, different environmental states
are orthogonal to each other (they can discriminate between different
states of the apparatus).
Equation~(\ref{deco}) is identical to the
density matrix of an ensemble of measurement results
$|n\rangle |\Phi_n\rangle$. System and apparatus thus seem to
be in one of the states $|n\rangle$ and $|\Phi_n\rangle$, given by
the probability $\vert c_n\vert^2$. This is decoherence.
The interaction with the environment singles out a preferred basis for
the apparatus, which is just the above-mentioned pointer basis.  

Assuming the usual Born rule for quantum probabilities, the reduced
density matrix contains all the information that can be obtained from
system and apparatus alone, without taking into account the quantum
entanglement with the environment explicitly. 

In our case it is the primordial fluctuations for which a classical
behaviour has to be obtained. Formally, they correspond to the
apparatus states above, but since we have no additional system, we
just call them the system variables; they are in interaction with the
environment. The latter is represented by other fields or by
higher-order modes of the metric and inflaton itself.  
 
An important property of a pointer basis is its preservation in time
-- they should be robust while interacting with the environment, that
is, they should not evolve into superpositions of themselves.
This basis thus represents the properties that become classical.
For this, two conditions have to be
fulfilled \cite{deco}: first, 
the projection operators on the pointer states 
should approximately commute
with the interaction Hamiltonian between system and environment;
second, these projection operators should approximately commute at
different times (in quantum optics, this is referred to as a quantum
non-demolition condition). We shall see in the following subsections
that these conditions are fulfilled 
in the high-squeezing limit for the field-amplitude basis of
the primordial fluctuations. 

In the literature, various methods are discussed with which one can
determine the pointer basis \cite{deco}. It has been shown that they
are equivalent to each other
at least in simple situations \cite{DK1}. One prominent
method makes use of the entropy of the system. As long as it is in a
pure state, this entropy is zero. If it is in a mixed state due to
decoherence, this entropy is positive. Its value characterizes the
degree of entanglement between system and environment -- the stronger
the entanglement, the higher the entropy. One can thus find the most
robust system states (that is, the pointer states) by minimizing the
resulting entanglement entropy. This method is called the
`predictability sieve' \cite{ZHP}. We shall apply this method below 
to the primordial fluctuations. 

A final tool for the study of the quantum-to-classical transition
should be introduced -- the Wigner function. For a general density
matrix in the position representation, $\rho(x,x')$, it is defined by
the expression
\be
\lb{Wigner}
W(x,p) = \frac{1}{\pi}\int_{-\infty}^{\infty} dy\ {\mathrm e}^{2ipy}
\rho(x-y,x+y) \ .
\ee
It is thus defined on phase space where it gives the correlations
between position and momentum. It is not, in general, positive
definite, reflecting the fact that quantum theory is inequivalent to a
classical statistical theory on phase space.

\subsection{ Wigner function and reduced density matrix for the
  primordial fluctuations}

We shall first address the system of primordial fluctuations as if it
were isolated \cite{PS1}. In order to recognize correlations between
position (here: field amplitude) and momentum (here: the variable
canonically conjugate to the field amplitude), the Wigner function can
be calculated.
In general it can assume negative values, but for the special case
of a Gaussian wave function it is everywhere positive (of a Gaussian
form).
This is the case here where the fluctuations are assumed to be in
their ground state. 
A convenient quantity is then the `Wigner ellipse' defined by the
contour of this Gaussian Wigner function, because it provides one directly
with a measure for the mean square deviations of $y$ and $p$.
 From the analysis of the Wigner function one can see that
the system behaves {\em as if}
it had a stochastic amplitude (with a Gaussian distribution if the
state is (\ref{psi})) and a fixed phase, see 
the calculations in \cite{PS1,KPS1,KP}.
This phase is given by the squeezing angle. The situation here is thus
already a peculiar one since the quantum nature of the squeezed state
cannot be seen, in the high-squeezing limit, from expectation values
and variances \cite{PS1}.

The scenario for the isolated system of primordial fluctuations
proceeds as follows. After the first horizon crossing (during
inflation) one has $r\lesssim 100$, $\varphi\to 0$, and the Wigner
ellipse will have -- after an appropriate re-scaling of the axes
-- a major half axis $\alpha=\exp(r)$ and a minor half axis
$\beta=\exp(-r)$. After the second horizon crossing (in the
radiation or matter dominated phase) the Wigner ellipse rotates,
with small oscillations around a large mean value of $r$. The
rotation is very slow, the frequency being about the inverse age
of the universe, which is why the squeezed nature is preserved for
a long time \cite{KLPS}.\footnote{In contrast to this, the
rotation is fast in the case of black holes that has important
consequences for Hawking radiation \cite{CK01}.} For modes that
re-enter in the radiation era, this rotation leads to the acoustic
peaks in the cosmic background radiation \cite{PS1,AFJP}. These
peaks have been observed with high precision \cite{Archeops,WMAP}.

This discussion treats fluctuations as being isolated, which 
is, however, unrealistic \cite{deco,Sakagami,BLM}.
Other fields (the `environment')
interact with them, even if the coupling is weak. Such a coupling
may also arise from a small self-interaction of the modes
\cite{LLN}. Even if there were no other fields present,
this small self-interaction would remain.
As long as the modes are outside the Hubble radius, this
coupling cannot lead to a direct causal interaction, but can only
produce entanglement (`EPR-type situation'). Since usual
interactions couple to fields (instead of canonical momentum), the
coupling is in $y$ (not $p$). This already suggests that the
$y$-basis is equal to the pointer basis at least approximately.
Spatial gradients would not change this conclusion, since they do not
depend on $p$.

The situation is analogous to the localization of particles in
quantum mechanics \cite{deco,JZ}. This interaction with the
environment without direct causal contact can be described by
multiplication of the density matrix $\rho_0(y,y')$ corresponding
to the system alone with a Gaussian factor according to \be
\lb{rhoxi} \rho_0(y,y')\longrightarrow
\rho_{\xi}(y,y')=\rho_0(y,y')
\exp\left(-\frac{\xi}{2}(y-y')^2\right)\ , \ee cf. \cite{KPS00,BLM}.
The parameter $\xi$ encodes phenomenologically the details of the
interaction strength with the environment. This interaction is by no
means restricted to be linear and can be very complicated.
In this way,
non-diagonal elements are suppressed with respect to the
$y$-basis. In a realistic situation for decoherence, one would
expect that $\xi$ dominates over the corresponding part in
\eqref{psi}, \be \lb{deccond} \xi\gg \Omega_{\rm R}\approx
k\E^{-2r} \ . \ee The typical time scale for decoherence during
inflation is $t_{\rm d}\sim H_{\rm I}^{-1}$. (There is a close analogy to
chaotic systems, with $H_{\rm I}$ corresponding to the Lyapunov parameter
\cite{KPS00}. The reason for this is the classical instability of
our system.)
 
 From analogous investigations in quantum mechanics \cite{ZP},
it can realistically be expected that the parameter $\xi$
settles to a constant value after a certain transition time.
The axes of the Wigner ellipse then read
\be
\alpha\approx \E^r\ , \quad \beta\approx\sqrt{\frac{\xi}{k}} \gg \E^{-r} \ ,
\ee
where (\ref{deccond}) has been used. While the major axis has remained
unchanged, the minor axis has become bigger and settles to a
constant value. The area of the Wigner ellipse thus increases, leading
to a non-vanishing entropy (see below).
In order to avoid conflict with observation, one
has to impose the `correlation condition' $\beta\ll\alpha$, leading to
\be
\lb{corrcond}
\frac{\xi}{k}\ll\E^{2r}\ .
\ee
Together with the slow rotation of the ellipse, this
guarantees the formation of acoustic oscillations.
A random distribution of $p$ and $q$ would totally smear out
these structures.

\subsection{Entropy}

As long as the fluctuations are treated as an isolated system, they can
be described by a wave function and thus have vanishing entropy.
This is no longer the case for the interacting system described by
(\ref{rhoxi}); the entanglement with the environment leads to a
`loss of information' for the system and therefore to a positive entropy.
The quantum correlations are present only in the total system
and are therefore unseen `locally'.
The entropy is calculated from
\be
\label{entropy}
S=-{\rm tr}(\rho_{\xi}\ln\rho_{\xi})\ ,
\ee
where we have set $k_{\rm B}=1$. 
The entropy has, of course, also to be calculated for a
two-mode squeezed state, although for simplicity we give
here again the calculation for one real mode only. For the full entropy
per the two-mode state,  one thus has to double the results below.

It is convenient to introduce the
dimensionless
parameter $\chi=\xi/\Omega_{\rm R}$ controlling the strength of decoherence.
 (In the case of pure exponential
inflation one has $\chi=\xi(1+4\sinh^2r)/k$.)
Inserting \eqref{rhoxi} into \eqref{entropy},
one gets the explicit expression
\be\label{S-exact}
\begin{split}
S&=-\ln\frac{2}{\sqrt{1+\chi}+1}-\frac12\left(\sqrt{1+\chi}-1\right)
                \ln\frac{\sqrt{1+\chi}-1}{\sqrt{1+\chi}+1} \\
 &=\ln\frac12\sqrt{\chi}
                -\sqrt{1+\chi}\ln\frac{\sqrt{1+\chi}-1}{\sqrt{\chi}}\ .
\end{split}
\ee
One recognizes that the entropy vanishes for $\xi\to 0$, as it must.
In the limit $\chi\gg1$ (large decoherence) one gets
\be
\lb{S-approx}
S=1-\ln2+\frac{\ln\chi}{2}+{\mathcal O}(\chi^{-1/2}) \ .
\ee
Both (\ref{S-exact}) and (\ref{S-approx}) are displayed in Figure~1.

\begin{figure}[htb]
\psfrag{chi}{$\chi$}
\psfrag{S}{$S$}
\psfrag{S-ex}[l][l]{$S$, \eqref{S-exact}}
\psfrag{S-dec}[l][l]{Approximation, \eqref{S-approx}}
\centering\includegraphics[width=\linewidth]{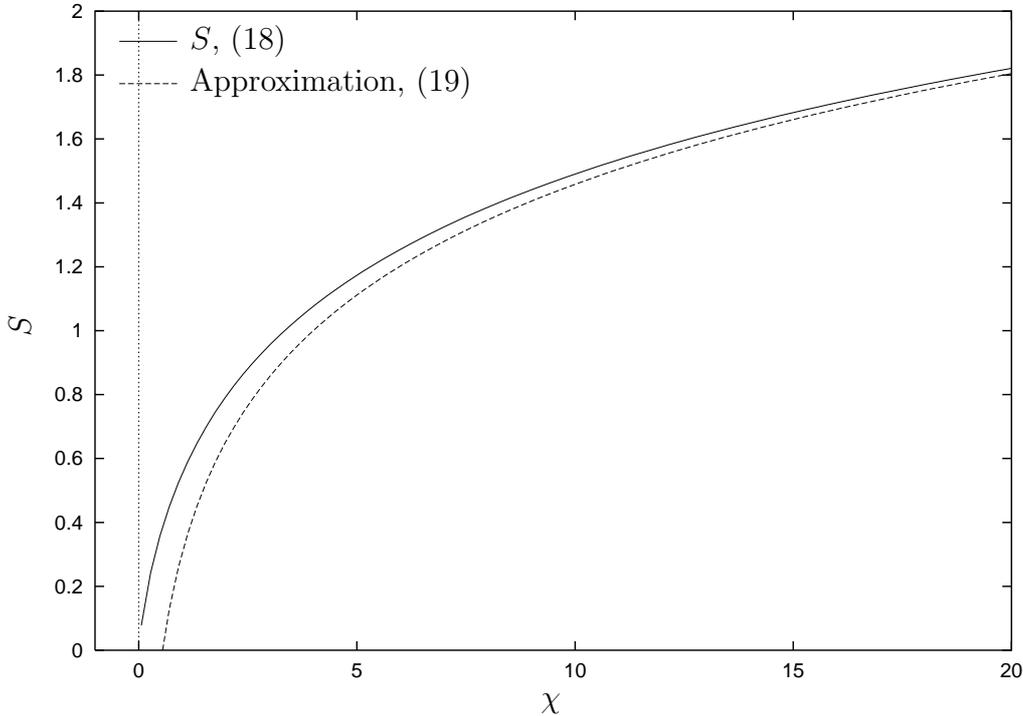}
\caption{Entropy in dependence on the decoherence parameter $\chi$.
Shown are the exact expression and the limiting case.}
\label{fig:S(chi)}
\end{figure}

It is seen that the asymptotic value is readily attained. For
$\chi\approx 1.5$ one obtains $S\approx\ln2$, corresponding to the
loss of one bit of information. Usually decoherence sets in
roughly at this stage \cite{deco}. A stronger condition than 
(\ref{corrcond}) is
obtained by the reasonable demand that some squeezing should
remain compared to the vacuum state which has $\Omega_{\rm R}=k$.
This leads to the condition $\xi<k$ and yields for exponential
inflation in the high-squeezing limit the bound $S\lesssim r$.
Whether this really holds is, of course, a question about realistic
interactions in the universe. 

The maximal entropy is instead given
by $S_{\rm max}=2r$ (obtained by smearing out the Wigner ellipse
into a big circle and corresponding to the pointer basis being the
particle-number basis \cite{prokopec}). One can thus define a notion
of {\em partial} decoherence in the entropy sense by the ratio of entropy to 
maximal entropy, $S/S_{\rm max}$, per each mode. 

It is also instructive to consider the linear entropy. It is defined by
\be
S_{\rm lin}={\rm tr}\left(\rho-\rho^2\right)\ ,
\ee
obeying $0\leq S_{\rm lin}<1$. Inserting (\ref{rhoxi}), one obtains
\be
S_{\rm lin}=1-(1+\chi)^{-1/2} \ .
\ee
For $\chi\ll1$ one gets $S_{\rm lin}\approx\chi/2$, while for
$\chi\gg1$ one gets $S_{\rm lin}\approx 1-\chi^{-1/2}<
1-\exp(-r)$, where we have used the bound $\xi<k$ in the last step.

As was already mentioned, the fact that the interaction with the
environment is in $y$-space leads to the suggestion that -- for
large squeezing -- the pointer basis equals approximately the
field-amplitude basis \cite{KPS1}. Exact diagonalization would
correspond to an entropy increase given by $S=r=S_{\rm max}/2$
\cite{KOZ}, in accordance with the expectation that $S<r$ holds
for modes outside the horizon. Contrary to this it was claimed in
\cite{CP1} that the pointer basis is given by (two-mode)
coherent states. Diagonalization in such a basis was in the
cosmological context first addressed in \cite{matacz}, with the
result that $S$ approaches for large squeezing the value $r=S_{\rm
max}/2$, that is, the same value as for the field-amplitude basis.
Figure~2 displays the entropy $S(r)$ for the field-amplitude
basis, the coherent-state basis (`$z$-basis'), and the
particle-number basis (`$n$-basis'). It is easily seen that $S$
approaches $r$ in the first two cases, whereas it approaches $2r$
in the latter case.

\begin{figure}[htbp]
\psfrag{r}{$r$}
\psfrag{S}[Bl][Bl]{$S$}
\psfrag{n: Red. in n-Basis}%
        {\scriptsize Reduction in $|n\rangle$-basis}
\psfrag{y: Red. in y-Basis}%
        {\scriptsize Reduction in $|y\rangle$-basis}
\psfrag{z: Red. in z-Basis}%
        {\scriptsize Reduction in $|z\rangle$-basis}
\psfrag{n}[cc][cc]{\textbf{n}}
\psfrag{y}[cc][cc]{\textbf{y}}
\psfrag{z}[cc][cc]{\textbf{z}}
\includegraphics[width=\linewidth]{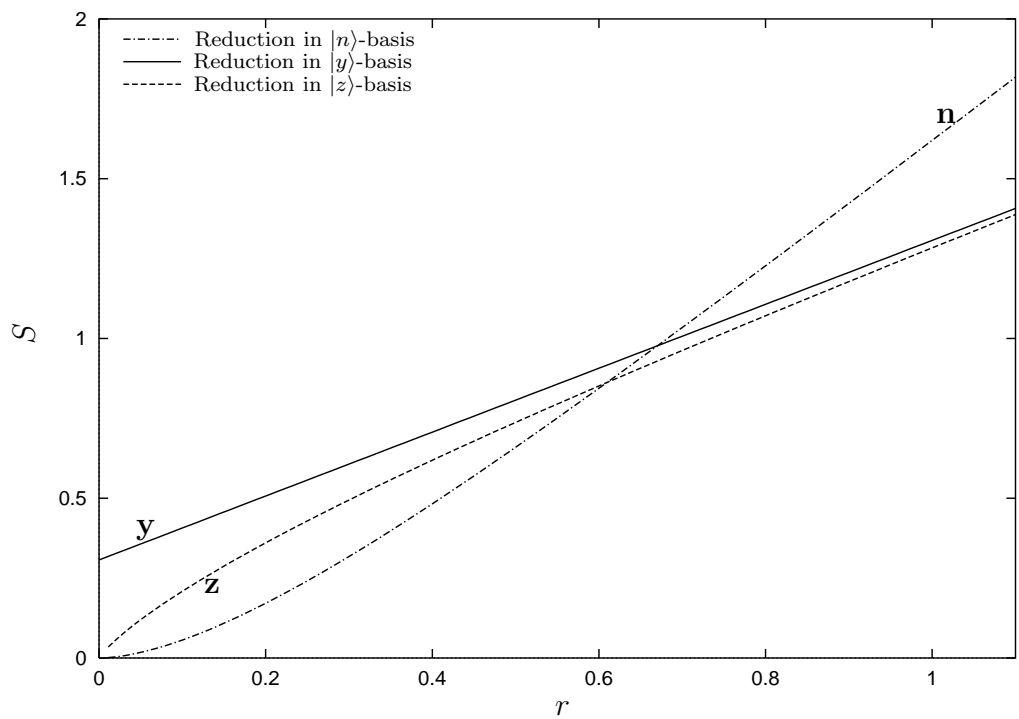}
\caption{Entropy $S$ in dependence of the squeezing parameter $r$
for three cases of the pointer basis.}
\label{fig:S(r)}
\end{figure}

\subsection{Pointer basis for the primordial fluctuations}

We shall now re-enforce from various points of view our earlier
result \cite{KPS1,KP} that the
pointer basis is for large squeezing
(approximately) given by the field-amplitude basis.

The first argument is about the normalizability of the pointer basis.
The authors in \cite{CP1} are concerned about the fact that
field-amplitude eigenstates are not normalizable and that they, moreover,
exhibit infinite momentum spread. However, even if one dealt with
exact amplitude (`position') eigenstates, the quantum mechanical formalism
would be able to cope with such a situation (GNS construction instead of
Hilbert space).
But the point is that the
environmental parameter $\xi$ entering (\ref{rhoxi}) is never strictly
infinite, so that a finite, albeit small, off-diagonal part for the
density matrix remains. One is thus dealing with very narrow
wave packets that are perfectly normalizable. 
Even if we had exact position eigenstates, an `infinite'
momentum spread would only take place if the kinetic term played a
crucial role in the dynamics. As has already been remarked above,
however, for the highly squeezed modes
the kinetic term is negligible compared to the
potential term.

This irrelevance of the kinetic term is also the reason why the reference
to \cite{ZHP} in \cite{CP1} is misleading. In \cite{ZHP}, the question of the
pointer basis has been studied for an ordinary harmonic oscillator --
in contrast to cosmology where one deals more with an upside down oscillator
\cite{GP} --
coupled to a high-temperature environment,
using the method of the `predictability sieve'
(minimal local entropy production). It was found that this basis is
given by the coherent-state basis.\footnote{This result had already been
found by another method -- implementing the rate with which
initially unentangled states become entangled -- in \cite{KZ}.}
A crucial ingredient in this proof,
however, was the importance of both the kinetic
and potential term for the oscillator (an averaging procedure
over many oscillation cycles had to be performed). For small times,
when the kinetic term does not yet become relevant, it was found in
\cite{ZHP} that, in fact,
the position basis is the (approximate) pointer basis.
This is the limit of relevance for the cosmological fluctuations.

In situations where the system Hamiltonian is
negligible, the pointer basis has the property that it consists of states
which are eigenstates of an operator that commutes with the interaction
Hamiltonian. If the system Hamiltonian is not negligible
(in our case, this happens for small $r$), one must invoke
a principle such as the `predictability sieve' or the `rate of 
de-separation' in order to determine a pointer basis.
In the case of a free particle coupled to a
localizing environment it was shown that the predictability sieve
(as well as other criteria) predict robust pointer states that are
narrow Gaussian wave packets \cite{DK1}. In the next section
we shall apply methods of open system dynamics to the Hamiltonian
(\ref{H}) in order to discuss the decoherence time for the
cosmological fluctuations with wavelengths both bigger
and smaller than the Hubble scale. This will enforce our earlier result 
about the nature of the pointer basis.

Another argument follows from the requirement that the pointer basis 
should be stable with respect to time evolution. In the limit of no 
interaction with environment, this means that it should be stable with 
respect to free evolution. However, the coherent state basis does not 
satisfy this requirement in the period between the two Hubble radius 
crossings: expansion of the Universe causes coherent states to become 
strongly squeezed ones in a characteristic time $\sim H^{-1}$.

This can be seen as follows.
It has been shown \cite{AFJP,SC} that the time evolution
generated by \emph{any} quadratic Hamiltonian like \eqref{H}
factorizes into an irrelevant phase rotation and the action of a
squeezing operator with parameters $r$ and $\varphi$.  It is further
well known (e.g.\ \cite{SZ} Sect.~2.7, and \cite{SC}) that a
squeezing operator acting on an \emph{arbitrary} coherent state
$|z\rangle$ (not just the vacuum ground state) produces a 
`squeezed coherent state': It has characteristic uncertainties
proportional to ${\rm e}^{-r}$ and ${\rm e}^r$ along rotated axes in the 
$y$-$p$ space, coinciding with the squeezed and the stretched semi-axes 
of the Wigner ellipse, respectively.
If we thus consider the time evolution of a
coherent state starting at a time $t_{\mathrm i}$ after the mode has
left the Hubble scale, we have to be concerned solely with the development
of the squeezing parameters $r$ and $\varphi$ starting with
$r=0=\varphi$ at that time.  As in the particular de~Sitter example
mentioned before, the squeezing angle $\varphi$ then quickly `freezes
out' and the squeezing factor grows like $r\approx \ln(a/a_\mathrm i)$, a
result first obtained for gravitational waves \cite{GS}, and later
generalized to scalar perturbations \cite{AFJP}. For
quasi-exponential inflation, we consequently have $r\approx
H_{\mathrm I} (t-t_\mathrm i)$, and it takes a very short time $\approx
H_{\mathrm I}^{-1}$ for the state to become significantly squeezed.
But even if we start in the ensuing radiation-dominated era, where $a
\propto (t/t_\mathrm e)^{1/2}$ and $H=1/(2t)$, we obtain $r\approx
\ln(t/t_\mathrm i)/2=\ln(2H_{\mathrm i} t)/2$, so that 
squeezing still becomes effective at a time $t\approx ({\rm e}^2/2) H_\mathrm
i^{-1}= {\rm e}^2 t_\mathrm i$ (with $H_{\mathrm i}$ denoting the
Hubble parameter at 
the initial time $t_\mathrm i$).  Finally, these considerations also
apply to two-mode coherent states (cf.\ the above references).
On the other hand, the field-amplitude basis is stable with respect to 
time evolution in the super-Hubble regime, since the field amplitude
remains constant with great accuracy.
Here, the squeezing manifests
itself in the continuous decrease of the decaying mode (see e.g.
\cite{DP99}).

An interesting and subtle situation arises with the second order adiabatic
basis proposed in \cite{amm05}. In the case of the exactly massless and
minimally coupled scalar field (\ref{H}) {\em and} the exact de Sitter
background, the vacuum mode functions $\tilde\phi_k$ of the second order
adiabatic vacuum (as defined in this paper) coincide with the exact 
solution for $\phi_k=y_k/a$. This has led the
authors of \cite{amm05} to the statement that there is no creation of
massless minimally coupled particles in the de Sitter space-time. However,
this result is a purely academic one since in case of the stable, eternal de
Sitter expansion there is no possibility for an observer to measure these
particles because they are beyond his/her future event horizon.\footnote{So,
it is reminiscent of the problem,  much discussed in the past, of radiation
reaction for a charged particle eternally moving in a constant electric
field.} On the other hand, for a viable inflationary model in which the
Hubble parameter $H$ decreases during inflation, the mode functions
$\tilde\phi_k$ {\em do not} coincide with the exact solution for $\phi_k$ 
which is constant far outside the Hubble radius ($k\ll aH$) and equal to 
$H(t_k)/\sqrt{2k^3}$ there ($t_k$ is the moment of the first Hubble radius 
crossing when $k=aH$). 

Really, the second-order adiabatic vacuum basis functions were defined in
\cite{amm05} in the form
\be
\label{adia}
\tilde\phi_k={1\over \sqrt{2a^3W_k}}\exp\left(-i\int^t dt'\,W_k(t')
\right)~,~~~ W_k={k\over a}\cdot {2k^2\over 2k^2+a^2(\dot H + 2H^2)}
\ee
for a spatially flat FRW model with an arbitarty $a(t)$. For a 
quasi-de Sitter (slow-roll) regime $|\dot H|\ll H^2,~a(\eta)\approx
-1/H(\eta)\eta~(\eta < 0)$, this expression reduces to
\be
\label{qDS}
\tilde\phi_k={H(\eta)\over \sqrt{2k}}e^{-ik\eta}
\left(-\eta +{i\over k}\right)~.
\ee
Outside the Hubble radius, $|\tilde\phi_k|=H(\eta)/\sqrt{2k^3}\propto H(t)$. 
Since $H(t)< H(t_k)$ for $t>t_k$, this means that creation of massless 
minimally coupled particles during slow-roll inflation does occur even 
with respect to this basis. By the end of inflation
and subsequent heating of matter, we return to standard formulas for the
particle number and the energy density power spectrum presented in
\cite{St79} (for the case of gravitons). Thus, in contrast to the 
field-amplitude basis, the second-order adiabatic basis (\ref{adia}) 
is also not 
stable with respect to time evolution in the super-Hubble regime even 
during an inflationary stage with slow-roll.

Of course, there exists an ambiguity in the definition of adiabatic 
vacuum mode functions of a given order (in other words, in the definition 
of the notion of number of particles in an external varying gravitational 
field) -- alternative definitions may differ by terms of a higher order in 
the adiabatic parameter $aH/k$ and its time derivatives. However, 
different variants of adiabatic mode expansion produce essentially the 
same form of the adiabatic expansion of the average value of the quantum 
field $\phi$ energy-momentum tensor as was shown already in \cite{ZS72,PF74}. 
Note in this connection that the most standard textbook calculation of a 
next small correction to the WKB solution of the wave equation
\be
\label{eweq}
\ddot\phi_k + 3H(t)\dot\phi_k + {k^2\over a^2(t)}\phi_k = 0
\ee
(that corresponds to the Hamiltonian (\ref{H})) in the WKB (sub-horizon) 
regime $k\gg aH$) reads
\be
\label{adia1}
\phi_k={1\over a\sqrt{2k}}e^{-ik\eta}(1+f)~,~~~f={i\over 2k}
\int dt\, a(\dot H + 2H^2)~,~~~|f|\ll 1~.
\ee
If we formally use (\ref{adia1}) for all $k$ during slow-roll inflation
ignoring the condition $|f|\ll 1$, it just reduces to the same expression 
(\ref{qDS}), exact for $H=const$ but significantly different from the exact 
solution for a slowly variable $H$. See also \cite{MS03,CFLV05,CFKLV06} for 
recent further improvements of the WKB-expansion in this case, in particular,
making it homogeneous in both the sub- and super-Hubble regimes. 

We would finally like to address the general criticism of
\cite{PSS} where it is claimed that in the discussion of emerging
classical behaviour for the fluctuations a transition from a symmetric
quantum state to a 
non-symmetric classical state is implicitly assumed without
justification. The authors of \cite{PSS} therefore claim that in
addition new physics (an explicit wave function collapse) is
needed. This is, however, a general issue in quantum theory and plays
a role, for example, in spontaneous symmetry breaking, see
\cite{Zehbook}, Sec.~6.1: the initial symmetric state  develops into a
symmetric superposition of all `false vacua'. 
The standard false vacuum is obtained by selecting one component out of this
superposition that can be justified using {\em any} interpretation
of quantum mechanics, for example the Everett or Copenhagen one, without
changing of quantitative predictions of quantum mechanics referring to
this component.
The same happens
here: the initial symmetric vacuum state evolves into a symmetric
superposition of inhomogeneous states out of which one component is
`selected' \cite{Zehbook}. Thus, cosmological perturbations are not
specific in this sense and 
neither solve nor complicate the fundamental problem of the foundations
of quantum mechanics.


\section{Decoherence time and pointer states}

\subsection{Description of decoherence by master equations}

Decoherence is the irreversible emergence of classical properties
for a quantum system through its unavoidable interaction with the
`environment', that is, with irrelevant degrees of freedom
\cite{deco}. Ideally, one would solve the Schr\"o\-dinger equation
for system plus environment and then trace out the environmental
degrees of freedom to obtain the reduced density matrix for the
system. This density matrix obeys a master equation which provides
a non-unitary and irreversible time evolution. Instead of
performing this procedure explicitly, one can directly make a
general ansatz for the master equation that can cope with all
interesting situations. A convenient form for this equation is the
`Lindblad form', cf. Sect.~3.3.2.2 in \cite{deco} or Eq. (7) in
\cite{BA}. The corresponding master equation is Markovian (local
in time) and preserves the properties of a density matrix (such as
conservation of its trace). Such an equation results from a wide
range of realistic interactions; the details of these interactions
are encoded in the Lindblad operators. The simplest form is a pure
localization term in addition to the system Hamiltonian; such a
term leads to the Gaussian suppression of interferences described
by (\ref{rhoxi}). The most general interaction may be non-Markovian,
but since we are interested in the minimal mechanism to guarantee
decoherence, the restriction to the Markovian case is sufficient. 

In \cite{DK2}, a master equation for the reduced density operator
$\hat{\rho}$ was studied for a free particle plus
such a localization term. It reads
\be
\lb{master}
\frac{\D\hat{\rho}}{\D t}=-\frac{\I}{2m\hbar}[\hat{p}^2,\hat{\rho}]
-\frac{D}{2}[\hat{x},[\hat{x},\hat{\rho}(t)]]\ .
\ee
Here, $D$ contains the strength of the interaction with the
environment and describes the strength of localization. The term
in \eqref{master} containing $D$ can arise from a high-temperature environment,
as it was discussed for example in \cite{ZHP}, but it does not have to.
It actually follows from a much wider class of situations \cite{JZ}.
Even if the environment is thermal, the temperature does not need
to be high; in a well-known example calculated in \cite{JZ} a
small dust particle in intergalactic space is localized
(decohered) by the ubiquitous microwave background radiation, a process
described by an equation of the form \eqref{master}.
This master equation is also obeyed by the density matrix for
the primordial fluctuations, which is discussed in \cite{KPS00}.

We shall here consider the situation where a general initial state 
(not necessarily Gaussian) is
present. In such a situation the Wigner function is usually not
positive definite (cf. \cite{WTFS} for the realization of such states
in the laboratory). The emergence of positivity is usually connected
with decoherence, so it can be used alternatively to the density
matrix as a measure for classicality.
It was found in \cite{DK2} that the Wigner function
for the case \eqref{master} becomes positive
after a certain time $t_{\rm d}$, independent of the initial state.
This, therefore, signals the emergence of classical behaviour.
Moreover, it was shown there that the reduced density operator
$\hat{\rho}$ can for $t>t_{\rm d}$ be decomposed in the form,
\be
\lb{pointer}
\hat{\rho}=\int\D\Gamma\ P(\Gamma,t)\vert\Gamma\rangle\langle\Gamma\vert
\ , \quad P(\Gamma,t)\geq0\ ,
\ee
where $\vert\Gamma\rangle$ denotes a set of Gaussian states.
They play the role
of the pointer states, so \eqref{pointer} is not the standard
orthogonalization of the density matrix;
note that the decomposition (\ref{pointer}) is with respect to an
overcomplete basis.
Localization leads here to the emergence of
narrow wave packets in position space (but not to delta functions).
The width of these narrow Gaussians was determined by the
`predictability sieve' \cite{DK1}. Apart from numerical factors of order 
one, the width is given by the expression $\delta\equiv(\hbar/mD)^{1/4}$.
If the kinetic term in \eqref{master} is neglected compared to
the localization term, we have $m\to\infty$ and thus
$\delta\to 0$ -- the width of the Gaussian pointer states becomes
very small, but they remain of course normalizable. For a small kinetic
term the pointer states are thus approximate position eigenstates.
This is the situation that has its analogue in the long-wavelength
modes in cosmology. The limit $m\to\infty$ is analogous to the limit
$r\to\infty$. This can be seen from the correlation 
$p\approx(\tan \varphi)y\approx {\rm e}^{-r}y$, which holds for
exponential inflation and expresses the smallness of the kinetic
($p^2$-containing) term.

We shall address below the question of positivity for the
Wigner function and the ensuing decoherence time
in the cosmological case. While for the special
initial state (\ref{psi}) the Wigner function is always positive,
the question of positivity is non-trivial for all other
initial states, such as the ones in \cite{LPS} (coherent states, states 
with an arbitrary number $N$ of particles).

The results of \cite{DK2} were generalized in \cite{BA} to
a system Hamiltonian of {\em arbitrary quadratic form} and to a general
Lindblad equation. It was shown in particular that the Wigner function
has always (apart from the case of Gaussian states) negative parts for
$t<t_{\rm d}$. The emergence of positivity at time $t_{\rm d}$
(and not earlier) is
thus independent of the initial state. It was generalized to
non-Markovian situations in \cite{Eisert}. It can thus be
assumed that in a generic situation of a particle coupled to an environment,
the state of the system can after a finite time not be distinguished
from an exact mixture of particular Gaussian states which have a narrow width
in case of strong interaction. 

\subsection{ Lindblad equation for the primordial fluctuations}

We shall restrict ourselves in the following to the case of one
Lindblad operator $\hat{L}$. The generalization to several Lindblad
operators is straightforward \cite{deco}. Such a generalization
is needed if one wants to accommodate non-linear effects by the
primordial modes themselves, that is, to express the influence
of `environmental' $k$-modes on `system' $k$-modes.
Such a master equation was used in \cite{Martineau,BHH}, whereby
the `system modes' were chosen to be the long-wavelength (observable)
modes and the `environmental modes' were chosen to be
the short-wavelength (unobservable) modes with wavelengths smaller or 
of the order of the Hubble radius.

The master equation then reads \cite{deco}
\be
\lb{Lindblad}
\frac{\D\hat{\rho}}{\D t}=-\I[\hat{H},\hat{\rho}]
+\hat{L}\hat{\rho}\hat{L}^{\dagger}-\frac12\hat{L}^{\dagger}\hat{L}\hat{\rho}
-\frac12\hat{\rho}\hat{L}^{\dagger}\hat{L}\ .
\ee
We further assume as in \cite{BA} that the Lindblad operator
is linear in our variables $p$ and $y$. The general system Hamiltonian
discussed in \cite{BA} is of the form
\be
\lb{H2}
\hat{H}=H_{11}p^2+2H_{12}py+H_{22}y^2\ .
\ee
To conform with the conventions used there, we will from now on work
with a slightly different set of canonical variables, defined by the
substitutions $p\to\sqrt{k}\, p$ and $y\to y/\sqrt{k}$ in \eqref{H}, so that
the Hamiltonian reads
\be
\lb{newH}
\hat H=\frac{k}2 \left(p^2+y^2\right)+\frac{a'}{a}yp\ .
\ee
In our
case we thus have $H_{11}=H_{22}=k/2$, and $H_{12}=a'/2a\equiv\cH/2$,
where $\cH\equiv aH$ denotes the conformal Hubble parameter. In the
analysis of \cite{BA}, the determinant of the quadratic form defined by
(\ref{H2}) plays a crucial role.
(We note that this analysis relies on a
Hamiltonian that does not explicitly depend on time. Strictly speaking,
therefore, our
treatment is only valid for times that are reasonably short
compared to the characteristic timescale given by $\cH$.)
More precisely, the sign of this determinant decides about the qualitative 
features of the decoherence time. In our case, this determinant is positive 
for $k>aH\equiv\cH$ (`elliptic case') and negative for $k<aH$ (`hyperbolic 
case'), that is, the change of sign just happens when the modes cross the 
Hubble radius.  For modes outside the Hubble radius, the hyperbolic case is 
of relevance. This turns out to be of crucial importance.
The `inverted oscillator' is the
paradigmatic example for the hyperbolic case \cite{GP}.

Further following the notation used in \cite{BA}, we write the Lindblad
operator in the form
\be L=(\bvec l'+\mathrm i\bvec l'')\cdot
\begin{pmatrix}p \\ y\end{pmatrix},\ee
where the real components of the `vectors'
\be
\bvec l'=\begin{pmatrix}\lambda' \\ \mu'\end{pmatrix}
\quad\text{and}\quad
\bvec l''=\begin{pmatrix}\lambda'' \\ \mu''\end{pmatrix}
\ee
contain all information on the interaction of the system with its
environment. We also introduce the `dissipation coefficient'
$\alpha\equiv\mu'\lambda''-\mu''\lambda'$ as well as the quantity
$\sigma\equiv 2\sqrt{-\det \bvec H}=\sqrt{\cH^2-k^2}$.
Note that $\sigma$ is real for $k\leq\cH$ (outside the Hubble radius)
and imaginary for $k>\cH$ (inside the Hubble radius).
The parameter $\alpha$ is non-vanishing if there is an energy
exchange between system and environment.
The real part of $\sigma$ is called `Lyapunov exponent' in \cite{BA},
which is in accordance with our discussion of the entropy in \cite{KPS00}.
It is non-vanishing only in the hyperbolic case.
The quantum-mechanical example \eqref{master} is recovered from
\eqref{Lindblad} in the special case of $\lambda'=\lambda''=0$, and one
can then identify $D=\mu'^2+\mu''^2$. This leads in particular to
vanishing friction, $\alpha=0$.

A quantitative result of \cite{BA}, following \cite{DK2},
 is that the positivity time,
$\eta_\mathrm p$, for the Wigner function is the solution of
\be \lb{etaP} \det\bvec M(-\eta)=\frac14\ , \ee
with a certain (time-dependent) matrix $\bvec M$ that appears in the 
solution for the Wigner function and that depends on the Lindblad 
operators. For the case of primordial fluctuations we obtain
\begin{multline}
\lb{M}
\det\bvec M(-\eta)\\
        =\frac1{(4\sigma k)^2}\left[
        \frac{\E^{4\alpha\eta}-2\E^{2\alpha\eta}
\cosh2\sigma\eta+1}{\alpha^2-\sigma^2}
                        A_{11}A_{22}
-\frac{\left(\E^{2\alpha\eta}-1\right)^2}{\alpha^2}A_{12}A_{21}
        \right],
\end{multline}
where the $A_{ij}$ denote the components of a (complex, symmetric) matrix
$\bvec A$ that depends on the components
of the Lindblad operator as well as on the system parameters
$\cH$ and $k$. For the reader's convenience we give here the explicit
expressions of the matrix elements, 
\begin{eqnarray*}
A_{11}&=&k^2\left(\lambda'-\frac{k\mu'}{\cH-\sigma}\right)^2
+k^2\left(\lambda''-\frac{k\mu''}{\cH-\sigma}\right)^2\ , \\
A_{22}&=&k^2\left(\lambda'-\frac{k\mu'}{\cH+\sigma}\right)^2
+k^2\left(\lambda''-\frac{k\mu''}{\cH+\sigma}\right)^2\ ,\\
A_{12}&=&k^2\left(\lambda'^2+\lambda''^2+\mu'^2+\mu''^2\right)
-2k\cH\left(\lambda'\mu'+\lambda''\mu''\right)=A_{21} \ .
\end{eqnarray*}
While
\bea
A_{12}A_{21}&=&\left[k^2\lambda'^2-2k\cH\lambda'\mu'+k^2\mu'^2\right.
\nonumber\\
& & \ + \left.k^2\lambda''^2-2k\cH\lambda''\mu''+k^2\mu''^2\right]^2
\eea
is real and positive for a general non-vanishing Lindblad operator,
$A_{11}A_{22}=A_{12}A_{21}+\left(2\alpha k\sigma\right)^2$ is real, but one
cannot determine its sign easily for modes inside the horizon, where
$\sigma^2<0$.

\subsection{Modes outside the Hubble radius}

Let us first treat a mode $k$ of the cosmological perturbations
{\em outside} the
Hubble radius ($\sigma^2>0$), that is, 
before the second Hubble-radius crossing. Since
the coupling to the environment can only lead to entanglement
and not to disturbance in that
situation, we take the dissipation coefficient $\alpha>0$ to be very small,
that is, $\alpha\ll\abs{\sigma}$. 
We note that for such modes,
\bdm
\sigma\approx\cH\left(1-\frac{k^2}{2\cH^2}\right)
\stackrel{k/\cH\to 0}{\longrightarrow} \cH \ .
\edm
Since we have for inflation,
\bdm
a(\eta)=-\frac{1}{(\eta-2\eta_{\rm e})H_{\rm I}}\ , 
\edm
we get $\sigma\eta\approx 2\cH\eta_{\rm e}-1\to 1$ at the end of inflation.
Then, the condition \eqref{etaP} gives, using $\alpha\ll\sigma$,
\bdm
1\approx \frac{A_{12}^2}{k^2\sigma^2}\left(\frac{\cosh 2\sigma\eta-1}
{2\sigma^2}-\eta^2\right)\ ,
\edm
yielding $\vert\eta_{\rm p}\vert\sim\sqrt{k/\vert A_{12}\vert}$. 
Choosing $\lambda''=0=\mu''$ in order to implement $\alpha=0$, we have
\bdm
A_{12}\approx k^2(\lambda'-\mu')^2+2\lambda'\mu'k(k-\cH)
\approx -2\lambda'\mu'k\cH\ ,
\edm
where in the last step we have assumed that the term
with $k\cH$ dominates. This then gives $\vert\eta_{\rm p}\vert\sim
\vert\lambda'\mu'\vert^{-1}$ and, consequently, for the
corresponding positivity time:
\be
\lb{tP}
t_{\rm p}\sim H_{\rm I}^{-1}\ln\frac{H_{\rm I}^{-1}\vert\lambda'\mu'\vert}
{a_0}\ .
\ee
At this moment, the entropy reaches the value 
\begin{displaymath}
S\approx H_{\rm I}t_{\rm p}\approx \ln\frac{H_{\rm I}^{-1}
\vert\lambda'\mu'\vert}{a_0}\ .
\end{displaymath}
Note that the quantity
$t_{\rm L}\equiv a_0/\vert\lambda'\mu'\vert$ is invariant under arbitrary 
rescaling of spatial coordinates and has the dimension of time.

The most striking feature in this result is its approximate
independence of the coupling parameters (they enter $t_\mathrm p$ only
logarithmically). In \eqref{rhoxi} the influence of the environment was
modelled by the parameter $\xi$, effectively describing localization in 
the field-amplitude basis as one special case of interaction. So, the 
result just derived supports the earlier claim in \cite{KPS00} that the 
details of the coupling of primordial fluctuations to the environment are 
not important. The independence of the details on the coupling is a 
general feature of systems characterized by a `Lyapunov exponent', 
independent of whether the system is chaotic or (as is the case here) 
just classically unstable \cite{ZP}.
The positivity time (\ref{tP}) is of the same order as the decoherence
time discussed in \cite{KPS1,KP,KPS00}. Both times are generally related
\cite{DK2}. In this limit the density matrix can thus be
decomposed into Gaussian pointer states according to \eqref{pointer}.
For strong coupling to the environment which leads to decoherence,
these pointer states are narrow Gaussians with respect to the
field amplitude. We note that concrete models in which a specific
interaction in the action is chosen lead to results in accordance with
our general expressions for decoherence time and entropy
\cite{LLN,Koks}. 

We also want to comment on the situation far outside the Hubble radius,
but in a radiation dominated universe. There the scale factor obeys
\be
a(\eta)=\frac{\eta}{H_{\rm I}\eta_{\rm e}^2}=a_{\rm e}\left(\frac{t}
{t_{\rm e}}\right)^{1/2}\ ,
\lb{rad}
\ee
and thus one has again $\sigma\eta\approx 1$. Repeating the above
calculations now yields for the decoherence time,
\be
t_{\rm p}\sim \frac{H_{\rm I}a_{\rm e}^2}{2(\lambda'\mu')^2}\ 
={H_{\rm I}t_{\rm L}^2\over 2}~.
\lb{tPrad}
\ee
For $H_{\rm I}t_{\rm L}\gg 1$, where here $t_{\rm L}= a_{\rm
  e}/\vert\lambda'\mu'\vert$, 
this is a much longer time than the positivity time during inflation,
which means 
that decoherence is here much less efficient than during inflation,
in accordance with our earlier result \cite{KPS00}.

The physical explanation of these results is the following. As was
emphasized in \cite{PS1}, omission of the decaying mode is sufficient
to get quantum decoherence. However, this mode should be really small 
to justify this procedure. In the super-Hubble limit, the only process 
making the decaying mode small is the expansion of the Universe;
interaction with the environment destroys the initial correlation between 
decaying and growing modes but typically makes the amplitude
of the decaying mode larger (that is reflected in the growth of the
entropy $S$). During inflation, the expansion of the Universe is much
faster (quasi-exponential) than during the radiation period (\ref{rad}). 
Thus, the decaying mode decreases much faster during inflation, which
explains the difference between (\ref{tP}) and (\ref{tPrad}). Also,
due to the exponential decrease of the decaying mode amplitude, 
coupling parameters describing real physical decoherence enter only 
logarithmically in $t_{\rm p}$. Note, however, that the fact that
$Ht_{\rm p}$ in the second case is larger does not mean that the
{\em amount of entropy} gained during decoherence is less in that
case, too. This requires special investigation.

It is also important that $Ht_{\rm p}\gtrsim 1$ in both cases. This means, 
in particular, that after the time period $\Delta t \sim t_{\rm p}$ 
after the first Hubble radius crossing, the characteristic {\em rms} 
amplitude of the decaying mode of $y$ is much less than the {\em rms} 
value of $y$ in the initial vacuum state. This holds because
the ratio of decaying to constant (`growing') mode is
$\exp(-3H(t-t_k))$, and thus this ratio becomes small even for a only
logarithmically large exponent.
 Therefore, the Wigner ellipse remains squeezed 
in one direction below its vacuum width even after the decoherence.  
This is another form of the result of Sec. 2 that $S < S_{\rm max}/2$.

\subsection{Modes inside the Hubble radius}

Let us now examine the situation {\em inside} the Hubble radius.
The great generality of
\eqref{Lindblad} makes it hard to read off the qualitative features of the
positivity time. (See, however, the argument in \cite{BA} that one only has
to analyze one special example of the elliptic system
in order to obtain its general
qualitative behaviour.) We want to consider a representative (and
realistic) environment that is also used in \cite{BA}: a thermal bath of
photons with average occupation number $n$. This calls for two Lindblad
operators defined as
\be
\begin{gathered}
\bvec l'_1=\begin{pmatrix}0\\ \sqrt{\alpha(n+1)}\end{pmatrix},
\quad
\bvec l''_1=\begin{pmatrix}\sqrt{\alpha(n+1)}\\ 0\end{pmatrix},\\
\bvec l'_2=\begin{pmatrix}0\\ \sqrt{\alpha n}\end{pmatrix},
\quad
\bvec l''_2=\begin{pmatrix}-\sqrt{\alpha n}\\ 0\end{pmatrix}.\\
\end{gathered}
\ee
The non-unitary part of \eqref{Lindblad},
which is given by the dissipation coefficient
$\alpha>0$, and the matrix $\bvec A$
then become a sum over the different Lindblad operators.
We then get
\be
A_{11}A_{22}=\left[(2n+1)2\alpha k\cH\right]^2
\quad\text{and}\quad
A_{12}A_{21}=\left[(2n+1)2\alpha k^2\right]^2\ .
\ee
The determinant \eqref{M} can now be written as
\begin{multline}
\det\bvec M(-\eta) \\
        =\frac{\left[(2n+1)k\right]^2}{4\abs{\sigma}^2}\left[
                \left(\E^{2\alpha\eta}-1\right)^2
                -\frac{\alpha^2}{\alpha^2+\abs{\sigma}^2}\frac{\cH^2}{k^2}
                \left(\E^{4\alpha\eta}-2\E^{2\alpha\eta}\cos2|\sigma|\eta+1
\right)
        \right]
\end{multline}
with $\abs{\sigma}^2=k^2-\cH^2$.
Again assuming the coupling to be small, $\alpha\ll\abs{\sigma}$, the
prefactor of the second term becomes much smaller than unity. If we exclude
the unrealistic case that
$\E^{4\alpha\eta}-2\E^{2\alpha\eta}\cos2|\sigma|\eta+1\gg
\left(\E^{2\alpha\eta}-1\right)^2$,
the second term can safely be neglected to give
\be
\det\bvec M(-\eta)\approx
\left[\frac{(2n+1)k}{2\abs{\sigma}}\left(\E^{2\alpha\eta}-1\right)\right]^2\ .
\ee
This yields a positivity time
\be
\label{positivity}
\eta_\mathrm p=
        \frac1{2\alpha}\ln\left(1+\frac{\abs{\sigma}}k\frac1{2n+1}\right)
        \stackrel{k\gg\cH}\approx
        \frac1{2\alpha}\ln\left(1+\frac1{2n+1}\right)\ ,
\ee
where the last approximation is valid for modes long after the second 
Hubble radius
crossing when $\abs{\sigma}\approx k$. (It can easily be checked
that the above line of thought as well as the result follow from
$\abs{\sigma}\approx k$ alone.) Using
Friedmann time, the positivity time \eqref{positivity} reads
\be
t_\mathrm p\approx \frac{a_{\rm e}^2}{16t_{\rm e}\alpha^2}
\ln^2\left(1+\frac1{2n+1}\right)\ .
\ee
The positivity time is independent of the
Hubble parameter $\cH$, but strongly depends
on the (small) coupling $\alpha$, in
sharp contrast to the situation outside the Hubble radius.
This is completely natural from the physical point of view since, inside
the Hubble radius, there is no division of modes into growing and
decaying ones: both linear independent solutions of the wave equation
corresponding to the Hamiltonian (\ref{H}) oscillate with amplitudes 
adiabatically decreasing with the expansion of the Universe; no more
squeezing occurs. Thus, in this regime, the Universe expansion may not
help us in getting decoherence, only real dissipative processes work
in this direction.  

It is interesting to note that the limit $n\to 0$ does \emph{not} lead 
to $t_\mathrm p/t_\mathrm e\to\infty$, but rather to 
$t_\mathrm p/t_\mathrm e\to (\ln^2 2)a_{\rm e}^2/16t_{\rm e}\alpha^2$.
This does not seem to be an artifact 
of the approximations: it is due to the vacuum energy of the photon bath, 
and may vanish after proper renormalization. 

Other concrete models of interaction may use an interaction of the form
$V=h^{ik}\phi_{,i}\phi_{,k}$ where $h^{ik}$ is a metric perturbation and 
plays the role of the system, while $\phi$ is an effective (`phonon') 
field describing small-scale perturbations of the thermal background and 
plays the role of the environment. Clearly, such an interaction results 
from the kinetic term of this field. 


\section{Conclusions and discussion}

Using methods from the quantum theory of open systems, we have shown 
that primordial fluctuations decohere when their wavelength becomes
much larger than the Hubble radius and that then the pointer basis is
given by narrow Gaussians which approximate the field amplitude basis. 
Consequently, the entropy per mode is smaller than half the maximal 
entropy. Thus, the density matrix remains squeezed in one direction as 
compared to the pure vacuum state. However,
this does not preclude the positivity of the Wigner function. If a 
particular interaction between the modes and other fields is given, one 
can calculate the Lindblad operators and all physical quantities 
in terms of this interaction and find the positivity time $t_{\rm p}$
after which the Wigner function becomes positive everywhere and, therefore, 
may be approximated by a classical distribution in the phase space.
In the super-Hubble regime, $t_{\rm p}$ explicitly depends on the Hubble 
time $H^{-1}$ and is during inflation only logarithmically larger than it.
On the other hand, in the sub-Hubble regime after the 
second Hubble radius crossing, $t_{\rm p}$ is independent of $H$ and
is totally determined by dissipation processes.
As for the width, this leads to an expression analogous to (but more 
complicated than) the width $(\hbar/mD)^{1/4}$ for the localization of 
a particle.

This crucial difference between decoherence mechanisms in the sub- and
super-Hubble regimes removes, we suppose, doubts regarding the 
possibility of (partial) decoherence during inflation expressed in the 
recent paper \cite{BHH}. The answer to the question in this paper why
we do not see similar decoherence in the Minkowski space-time is, first,
because there is no super-Hubble regime in this case. Second, it was
assumed in \cite{BHH} that the `environmental' (short-wavelength) modes 
are in their adiabatic ground state. Fields in 
their ground state are usually not able to exert a decohering
influence \cite{deco}. 
So, the analysis presented here assumes that some fields are present 
which are {\em not} in their ground state. The concrete examples of such
fields are cosmological perturbations themselves, scalar perturbations
and gravitational waves, with scales of the order of the Hubble radius
(the case not considered in \cite{BHH}). But even if this was not true,
the analysis in \cite{PS1} has shown that it is impossible in practice
to distinguish the squeezed state of the modes from a classical stochastic 
ensemble, cf. also the gedanken experiments discussed in \cite{KP,KLPS}.

At last, it should be emphasized that we do not consider decoherence of
{\em generic} long-wavelength modes but specifically only the decoherence
of very strongly squeezed states. Then an exceedingly small interaction
{\em is} already sufficient for a loss of quantum coherence, which has
$S\gtrsim 1$. This fact, namely that, even in the Minkowski space-time,
strongly squeezed states are much more fragile than, for example, 
coherent states, is well known in standard quantum mechanics and 
represents the main obstacle to generate strongly squeezed states in the 
laboratory, see for example Sect.~3.3.3.1 in \cite{deco}.

Still it should be noted that there is an agreement between our paper 
and \cite{BHH} regarding the pointer basis in the super-Hubble regime.
On the other hand, we would not fully support the statement of 
\cite{Martineau} that decoherence is ``extremely effective'' during
inflation. What follows from our results is that decoherence, though
quick and sufficient to reach the positivity of the Wigner function, 
is not sufficiently effective, for example, to make the Wigner ellipse
exceeding its vacuum value in all directions (the latter would
correspond to $S > S_{\rm max}/2$).\footnote{Note also that the ratio
of inflaton gravitational to self-interaction is overestimated in 
\cite{Martineau}. One should take into account that, during inflation,
the gravitational potential $\Phi$ describing scalar perturbations is 
not constant, but slowly growing from zero at the first Hubble radius 
crossing up to its final value after the end of quasi-exponential
expansion of the Universe. Thus, its value during inflation is much 
less than that after inflation which can be observed now. Proper
account of this fact makes the inflaton gravitational and 
self-interactions of the same order.} 
Thus, our results regarding the degree of decoherence reached during 
inflation are, in some sense, intermediate between those of 
\cite{Martineau} and \cite{BHH}. This discussion shows how subtle is
the problem of decoherence and quantum-to-classical transition
for cosmological perturbations.
 

\section*{Acknowledgements}

C.K. is grateful to the Max Planck Institute for Gravitational
Physics in Golm, Germany, for its kind hospitality while part of
this work was done. He also thanks Robert Brandenberger, David Lyth,
and Daniel Sudarsky for discussions.
A.A.S. was partially supported by the
Russian Foundation for Basic Research, grant 05-02-17450, and by
the Research Program ``Quantum Macrophysics" of the Russian Academy
of Sciences, as well as by the German Science Foundation under
grant 436 RUS 113/333/10-2. He also thanks the Centre Emile Borel,
Institut Henri Poincar\'e, Paris, for hospitality in the period when 
this paper was finished.


\end{document}